\documentclass[prl,aps,superscriptaddress,amsmath,amssymb,twocolumn]{revtex4-1}
\usepackage{braket,verbatim,bm,bbold,amsmath,amssymb,epsfig,float,color}
\usepackage[colorlinks,bookmarks=false,citecolor=blue,linkcolor=blue,hyperfootnotes=true,urlcolor=blue]{hyperref}

\newcommand{\bw}{\begin{widetext}}
\newcommand{\ew}{\end{widetext}}
\newcommand{\be}{\begin{equation}}
\newcommand{\ee}{\end{equation}}
\newcommand{\bea}{\begin{eqnarray}}
\newcommand{\eea}{\end{eqnarray}}

\definecolor{violet}{rgb}{0.62,0,1}
\definecolor{lightblue}{rgb}{0.12,0.56,1}
\definecolor{green}{rgb}{0.13,0.55,0.13}

\newcommand{\Tr}[1]{\mathrm{Tr} #1}

\newcommand{\modified}[1]{{#1}}

\begin{document}
	
	\begin{titlepage}

		\title{Universal Thermal Corrections to Symmetry-Resolved Entanglement Entropy and Full Counting Statistics}
		
		\author{Mostafa Ghasemi}
		\email{Ghasemi.mg@ipm.ir}
		\affiliation{School of Particles and Accelerators, Institute for Research in Fundamental Sciences (IPM)\\
			P.O. Box 19395-5531, Tehran, Iran}

		\begin{abstract}
			 We consider the symmetry-resolved R\'{e}nyi and entanglement entropies for two-dimensional conformal field theories on a circle at nonzero temperature. We assume a unique ground state with a nonzero mass gap induced by the system's finite size and then calculate the leading corrections to the contributions of individual charge sectors in a low-temperature expansion. Besides the size of the mass gap and the degeneracy of the first excited state, these universal corrections depend only on the four-point correlation function of the primary fields. We also obtain thermal corrections to the full counting statistics of the ground state and define the \textit{probability fluctuations} function. It scales as $e^{-2 \pi \Delta_{\psi} \beta /L}$, where $\Delta_{\psi}$ is the scaling dimension of the lowest weight states. As an example, we explicitly evaluate the thermal corrections to the symmetry-resolved entanglement entropy and FCS for the spinless fermions.
			
		\end{abstract}
		
		\maketitle
		
		
		
	\end{titlepage}

	\emph{Introduction--}
	One of the outstanding features of quantum mechanics is the concept of entanglement~\cite{Horodecki:2009zz}, which relates to non-local correlations between various subsystems of a given quantum system. It has a wide range of applications in a diverse area of physics. Its measures, especially the entanglement entropy (EE), can be used as an order parameter for characterizing phase transitions \cite{Vidal:2002rm,Amico:2007ag,
		Laflorencie:2015eck} and topological phases \cite{Kitaev:2005dm,Levin:2006LW} in many-body quantum systems; providing new tools for quantifying quantum correlations and describing renormalization group flow in quantum field theory \cite{Calabrese:2004eu,Calabrese:2009qy,Casini:2009sr,Casini:2006es,Casini:2012ei}; Entanglement entropy also
	plays an important role in 
	the connection between quantum states  and quantum gravity in the framework of holography
	\cite{Rangamani:2016dms, Nishioka:2018khk} and is a key concept in describing the physics of the black hole information loss paradox
	\cite{Bombelli:1986,Srednicki:1993im}.

	The computation of the EE often based on the replica trick, by introducing $n$ copies of the system. For bipartite quantum system $A\cup B$, the $n$th R\'{e}nyi entropy (RE) is defined as $S_n \equiv \frac{1}{1-n} \log {\mathrm{Tr}}(\rho_A)^n$, where $\rho_A = {\mathrm{Tr}}_B \rho$ is the reduced density matrix of subsystem $A$. EE is given by $S_E=-{\mathrm{Tr}} \rho_A \ln \rho_A = \lim_{n \to 1} S_n$.
	Based on the path integral language~\cite{Calabrese:2004eu,Calabrese:2009qy}, the calculation of  $S_n$ reduces to computing the partition function on a Riemann surface geometry $\mathcal{R}_n$. Using the above approach, the authors \cite{Goldstein:2017bua} via the insertion of an \modified{Aharonov-Bohm} flux in the Riemann geometry could decompose entanglement measures into 
	the contribution of individual charge sectors in the presence of global symmetry.
	Hence, the computation of the path integral in the presence of this flux gives the quantity, charge moments $\mathcal{Z}_n(\alpha) = {\mathrm{Tr}} \left( \rho_A^n e^{i \alpha \hat{Q}_A} \right)$, that enable us to identify and compute the contributions to the entanglement related to each symmetry sector. $\hat{Q}_{A}$ is the total charge in the $A$ subsystem.
	
	More generally, in the presence of global symmetry, the reduced density matrix can be decomposed into block diagonal form associated with each charge sector, $\rho_{A}=\oplus_{Q} P(Q_{A})\rho_{A}(Q_{A})$. $P(Q_{A})$ is the probability of finding $Q$ as an outcome of a measurement of $\hat{Q}_{A}$. Symmetry resolved entanglement entropy is defined as $S(Q)=-{\mathrm{Tr}}\rho_{A}(Q) \ln \rho_{A}(Q) $. Total entanglement entropy is the sum of the contributions of individual charge sectors,  $S_{E}=\sum_{Q_A}P(Q_A)S(Q_A)-\sum_{Q_A}P(Q_A)\log (P(Q_A))=S^{c}+S^{n}$.
	$S^{c}$ is the configurational entropy \cite{lukin:2018}, which measures the average of the entanglement in each charge sector. $S^{n}$ is called number entropy\cite{lukin:2018,wv-03,kusf2,bhd-18}, which quantify the entropy due to the fluctuations of the charge within subsystem $A$. Another interesting quantity that can be deduced from charged moments is the full counting statistics (FCS). FCS is a measure of the distribution of the charge in a given region. There is a relation between entanglement spectrum and the FCS of charge fluctuations between two subsystems \cite{Klich:2008un,Song:2010su,Song:2011gv,Calabrese:2011ycz,Bastianello:2018vxa,Calabrese:2020tci,Ares:2020opo,Groha:2018sci,Arzamasovs:2019sua}.   
	The authors \cite{Goldstein:2017bua} provide theoretical framework to evaluating the contribution of each symmetry sector by relating the symmetry-resolved entanglement entropy to
	the Fourier transform of partition function on the $n$-sheet Riemann surface with generalized Aharonov-Bohm flux: $ \mathcal{Z}_n(Q_A)=\Tr{\rho^n_A \mathcal{P}_{Q_A}}=\int_{-\pi}^{\pi}{\frac{\textrm{d}\alpha}{2\pi} {\mathcal{Z}}_n(\alpha)e^{-i \alpha Q_{A}}}$.
	$\mathcal{P}_{Q_A}$ denotes the projection into the subspace of states of region $A$ with charge $\hat{Q}_A$. Symmetry-resolved R\'{e}nyi and entanglement entropies can be obtained as : $S_{n}(Q_A) = \frac{1}{1-n}\log \left[ \frac{\mathcal{Z}_n(Q_A)}{\mathcal{Z}_1^{n}(Q_A)} \right]$ and $S(Q_A)=\lim_{n\rightarrow1} S_{n}(Q_A)$.
	Probability is then $P(Q_A)=\mathcal{Z}_1(Q_A)$.
	The other related works are 
	\cite{Laflorencie2014,Xavier:2018kqb,Cornfeld:2018sac, Cornfeld:2018wbg,Feldman:2019upn,Bonsignori:2019naz,Fraenkel:2019ykl,Murciano:2019wdl,Capizzi:2020jed,Murciano:2020lqq,Turkeshi:2020yxd,Murciano:2020vgh,Horvath:2020vzs,Azses:2020wfx,Bonsignori:2020laa,Parez:2020vsp,Estienne:2020txv,Zhao:2020qmn,Vitale:2021lds,Murciano:2021djk,Horvath:2021fks,Chen:2021pls,Capizzi:2021zga,Horvath:2021rjd,Parez:2021pgq,Weisenberger:2021eby,Capizzi:2021kys,Milekhin:2021lmq,Chen:2021nma,Oblak:2021nbj,Parez:2022xur,Ares:2022hdh,Zhao:2022wnp,Jones:2022tgp,Belin:2013uta}.
	
	It is well known that entanglement entropy is a good entanglement measure for quantum systems in their ground state. However, the real world not lying at a zero temperature regime, and hence, the entanglement entropy is no longer proper for thermal states.  At finite temperature, the entanglement entropy of subsystem $A$ is contaminated by thermal fluctuation and, in fact, in the high-temperature limit becomes dominated by thermal entropy. To determine the quantum entanglement of the thermal systems, one should subtract off the thermal contribution to entanglement entropy. For the systems with the mass gap $m_{\text{gap}}$, the authors \cite{Herzog:2012bw} conjectured that, in the limit $\beta m_{\text{gap}}\gg 1$, these corrections scale as $e^{-\beta m_{\text{gap}}}$.
	The authors \cite{Cardy:2014jwa} calculated the coefficient of the Boltzmann factor by putting the conformal field theory on the cylinder and introducing the mass gap between the ground state and the first excited state through the finite size of the system.  They found that these corrections are universal and depend on the size of the mass gap and the degeneracy of the first excited state.

	The interesting questions that motivated this work are: What are the thermal corrections to the contribution of individual system charge sectors? How are these corrections scale? If these corrections are universal, what are their physical meaning? In this work, we are addressing these questions. We introduce \textit{the thermal charged moments} and based on the approach \cite{Cardy:2014jwa}, we derive the low-temperature expansion of it. We find that these thermal corrections are encoded in the four-point function of primary fields, the scaling dimension of the lowest weight primary field, and its degeneracy. Consequently, we can find the thermal corrections to the symmetry-resolved R\'{e}nyi and entanglement entropies. 
	We also obtain thermal corrections to the full counting statistics of the ground state (FCS) and define the fluctuations of probabilities. It scales as $e^{-2 \pi \Delta_{\psi} \beta /L}$.
	
	\emph{ Thermal charged moments and universal corrections--}
	The main quantity for computing the symmetry-resolved entanglement entropy is charge moments $\mathcal{Z}_n(\alpha)$. So to calculate the thermal corrections, we introduce \textit{the thermal charged moments} and then find its low-temperature expansion. To do so, we
	start with the thermal density matrix $\rho = \frac{e^{-\beta H}}{\Tr{e^{-\beta H}}}$, where $\beta$ and  $H$ are the inverse temperature and the Hamiltonian, respectively.
	The thermal density matrix can be written as a Boltzmann sum via introducing a complete set of states: $\rho = \frac{1}{\Tr{e^{-\beta H}}} \sum_{|\phi \rangle} |\phi \rangle \langle \phi | \, e^{-\beta E_{\phi} }$ \cite{Cardy:2014jwa}.
	We consider a unitary CFT with central charge $c$ defined on an infinite cylinder of circumference $L$ with coordinate $w=y- i t$. The corresponding Hamiltonian is $H = \left( \frac{2\pi}{L} \right) \left(L_0 + \tilde L_0 - \frac{c}{12}\right)$,
	where $L_0$ and $\tilde L_0$ are the zeroth level left- and right-moving Virasoro generators, respectively. Next, we assume that the finite size of the system induces a unique ground state and 
	gaps the theory between the ground state and the first excited state.
	According to state-operator correspondence, for any Virasoro primary operator $\psi(w) \neq 1$ there is an state $|\psi \rangle = \lim_{t \to -\infty} \psi(w) |0 \rangle$ where $L_0 |\psi \rangle = h_{\psi} |\psi \rangle$, $\tilde L_0 | \psi \rangle = \tilde h_{\psi} | \psi \rangle$ and $\Delta_{\psi} = h_{\psi} + \tilde h_{\psi}$, and vice versa. 
	If the $|\psi \rangle$ are lowest weight states in the CFT, then the smallest nonzero eigenvalue $E_\psi = \frac{2 \pi}{L} \left( \Delta_{\psi} - \frac{c}{12} \right)$  must correspond to a primary operator $\psi(w)$.
	Note that when $\Delta > 2$, two descendants of the identity operator, the stress tensor $T(w)$ and its conjugate $\tilde T(\bar w)$, give the dominant correction \cite{Cardy:2014jwa}.
	
	The low-temperature expansion of thermal density matrix becomes as
	\be
	\label{rhoexpand-l1}
	\nonumber
	\rho \sim \rho_{0}+ e^{-2 \pi \Delta_{\psi} \beta /L}\left(\rho_{\psi} -\rho_{0} \right)
	\ ,
	\ee
	where $\rho_{0}= |0\rangle \langle 0 |$ and $\rho_{\psi}=| \psi \rangle \langle \psi |$ are density matrices associated to vacuum and first excited states, respectively.  By partitioning the spatial circle into regions $A$ and $B$, we can compute the reduced density matrix $\rho_A$ by tracing out the degrees of freedom inside the region $B$.
	Accordingly,
	the low-temperature expansion of reduced density matrix becomes as
	\be
	\label{rhoexpand-l2}
	\rho_{A} \sim \rho_{0,A}+ e^{-2 \pi \Delta_{\psi} \beta /L}\left(\rho_{\psi,A} -\rho_{0,A} \right),
	\ee
	where $\rho_{0,A}={\mathrm{Tr}}_B (|0\rangle \langle 0 |) $ and $\rho_{\psi,A}={\mathrm{Tr}}_B(|\psi \rangle \langle \psi |)$. The $\Delta_{\psi}$
	is the smallest scaling dimension of primary operator, $\psi\left(w\right)$, among the set of  non-identity  operators including the stress tensor. 
	We define \textit{the thermal charged moments} as
	\begin{eqnarray}
	\label{EQTH-1}
	{\mathcal{Z}}_n^{\text{(th)}}(\alpha) &=& {\mathrm{Tr}} \left( \rho_A^n e^{i \alpha \hat{Q}_A} \right) \nonumber \\
	&&
	={\mathcal{Z}}_n^{(0)}(\alpha)+n{\mathcal{Z}}_n^{(0)}(\alpha) e^{-2 \pi \Delta_{\psi} \beta /L}\mathcal{F}_{n}(\alpha) . \ 
	\end{eqnarray}
	The first term is nothing but the charged moments at zero temperature that gives the resolved-symmetry entanglement entropy for a finite system \be
	{\mathcal{Z}}_n^{(0)}(\alpha)= c_{n,\alpha} \left[ \frac{L}{\pi} \sin \left( \frac{ \pi \ell}{L} \right) \right]^{ \frac{1-n^2}{6n}-2\frac{\Delta_{\mathcal{V}}}{n}},
	\nonumber\\ 
	\ee
	where $c_{n,\alpha}$ is non-universal constant, $\ell$ is the length of $A$, and $\Delta_{\mathcal{V}}$ denotes the scaling dimension of the operator $\mathcal{V}$ generating the Aharonov-Bohm flux.
	$\mathcal{F}_{n}(\alpha)$ is defined as: 
	\bea\label{M-EQ4}
	&&	\mathcal{F}_{n}(\alpha)=\mathcal{M}_{n}(\alpha)-1,
	\cr \nonumber\\
	&&
	\mathcal{M}_{n}(\alpha)=\frac{ \Tr{\left( \rho_{0,A}^{n-1} \rho_{\psi,A} e^{i \alpha \hat{Q}_A}\right) }}{ \Tr{\left( \rho_{0,A}^{n} e^{i \alpha \hat{Q}_A}\right) }}.
	\eea
	The denominator is the charged moments at zero temperature. The nominator is a new term. It can be interpreted as a two-point function in the presence Aharanon-Bohm flux.
	The expression (\ref{M-EQ4}) then takes the following form,
	\be 
	\label{MA-FO}
	\mathcal{M}_{n}=\frac{\left\langle \mathcal{V_{\alpha}}\mathcal{V_{-\alpha}}\psi \psi\right\rangle_{\mathcal{R}_n}}{\left\langle \mathcal{V_{\alpha}}\mathcal{V_{-\alpha}}\right\rangle_{\mathcal{R}_n}  \left\langle \psi \psi\right\rangle_{\mathcal{R}_1}}.
	\ee
	The numerator term is a four-point function of two $\mathcal{V_{\alpha}}$ and two $\psi$ in the replicated geometry $\mathcal{R}_n$, which is the multisheeted cylinder, for a particular insertion of points. The denominator terms are two two-point functions of the $\mathcal{V_{\alpha}}$ in the replicated geometry  $\mathcal{R}_n$ and $\psi $ in the original geometry  $\mathcal{R}_1$, cylinder.
	According to the Riemann-Hurwitz theorem, this multisheeted cylinder has genus zero. Hence, through the uniformizing map which takes the multisheeted cylinder to the plane, we can evaluate these correlation functions on the plane. For details see \hyperref[appendix1]{SM}. 
	In any way, we find then that
	\be \label{EQ-M1}
	\mathcal{M}_{n}(\alpha)=\frac{1}{n^{2\Delta_{\psi}}}\frac{\sin ^{2\Delta_{\psi}}\left(\pi x\right)}{\sin ^{2\Delta_{\psi}}\left(\pi x/n\right)}G_{n,\alpha}(z,\bar{z}).
	\ee
	where $x=\frac{l}{L}$. The $G_{n,\alpha}(z,\bar{z})$ is a function of cross ratios $z=\frac{\zeta^{(n)}_{12}\zeta^{(n)}_{34}}{\zeta^{(n)}_{14}\zeta^{(n)}_{23}}$, $\bar{z}=\frac{\bar{\zeta}^{(n)}_{12}\bar{\zeta}^{(n)}_{34}}{\bar{\zeta}^{(n)}_{14}\bar{\zeta}^{(n)}_{23}}$. This is one of main results in this work. All information about the thermal corrections to the charged moments and the symmetry-resolved entanglement entropy, is encoded in the $G_{n,\alpha}(z,\bar{z})$, which has a conformal block expansion. It can be regarded as the building block of our computations.
	
	\emph{ Full counting statistics--}
	Thermal charged moments can be defined as a generating function of FCS at finite temperature. FCS defines the distribution probability of conserved charge in the subsystem $A$ with length $l$.
	It can be defined via generating function $\chi(\alpha)=\sum_{Q_{A}=-\infty}^{\infty} P(Q_{A}) e^{ i\alpha Q_{A}}=\left\langle  e^{ i\alpha \hat{Q}_{A}}\right \rangle $,
	which encodes all the cumulants $C_{m}$,
	$\ln \chi(\alpha)=
	\sum_{Q_{A}=1}^{\infty}\frac{(i \alpha)^{m}C_{m}}{m!}$, where $C_{m}=(-i\partial_{\alpha})^{m}\ln \chi(\alpha)$.
	$C_{m}$ describe properties of the distribution probability $P(Q_{A})$. For example, the mean $C_{1}=\left\langle  \hat{Q}_A \right\rangle $, the fluctuations $C_{2}=\left\langle  \left( \hat{Q}_A-\left\langle  \hat{Q}_A \right\rangle \right) ^{2} \right\rangle $, and so on.
	In this section, we derive the thermal corrections to the FCS.
	Let us first define the quantity:
	\be \label{UNIV-1}
	f_{n}(\alpha,T)=\frac{{\mathcal{Z}}_n^{\text{(th)}}(\alpha)}{{\mathcal{Z}}_n^{(0)}(\alpha) }= 1+n\mathcal{F}_{n}(\alpha)e^{-2 \pi \Delta_{\psi} \beta /L}.\nonumber\\
	\ee
	If we take a logarithm of the above expression, we reach the universal quantity $	g_{n}(\alpha,T) = \log f_{n}(\alpha,T)$ that can be used to define the excess-cumulant generating function, such that, its different derivative in $\alpha=0$ gives the excess of various moments of  $\hat{Q}_{A}$,
	\begin{eqnarray} 
	\label{dFCS-2}
	\Delta C_{n,m} &=& \left(- i \partial_{\alpha} \right)^{m}g_{n}(\alpha,T)\arrowvert_{\alpha=0} \nonumber \\
	&& =\frac{ge^{-2 \pi \Delta_{\psi} \beta /L}}{n^{2\Delta_{\psi}-1}} \frac{\sin ^{2\Delta_{\psi}}\left(\pi x\right)}{\sin ^{2\Delta_{\psi}}\left(\frac{\pi 
			x}{n}\right)}
	\left(- i \partial_{\alpha} \right)^{m} G_{n,\alpha}(z,\bar{z}) \arrowvert_{\alpha=0}.
	\end{eqnarray}
	The relation (\ref{dFCS-2}), for $\Delta C_{1,m}=\Delta (\Delta Q_A^{m})$ with  $\Delta Q_A = Q_A - \langle  \hat{Q}_A \rangle$, denotes that the exceed-FCS depend on the mass gap, degeneracy, and the field content of the theory. The symmetry-resolved thermal partition function becomes
	\bea\label{Zn-corr}
	&&	{\mathcal{Z}}_n^{\text{(th)}}(Q_A)={\mathcal{Z}}_n^{(0)}(Q_A)\left[1+gn \mathcal{F}_{n}(Q_A)e^{-2 \pi \Delta_{\psi} \beta /L} \right],
	\cr \nonumber\\
	&&
	\mathcal{F}_{n}(Q_A)=\frac{1}{n^{2\Delta_{\psi}}}\frac{ \sin^{2\Delta_{\psi}} \left(\pi x \right) }{\sin^{2\Delta_{\psi}}\left( \frac{\pi x }{n}\right)}\frac{\mathcal{X}_{n}(Q_A)}{{\mathcal{Z}}_n^{(0)}(Q_A)}-1,
	\cr\nonumber\\
	&&
	\mathcal{X}_{n}(Q_A)=\int_{-\pi}^{\pi}{\frac{\textrm{d}\alpha}{2\pi} {\mathcal{Z}}_n^{(0)}(\alpha)G_{n,\alpha}(z,\bar{z})e^{-i \alpha Q_{A}}}.
	\eea
	The leading correction term to the probability distribution of charge in each sector can be obtained by  $P^{\text{(th)}}(Q_A)={\mathcal{Z}}_1^{\text{(th)}}(Q_A)$. FCS was previously calculated for the ground state \cite{bss-07, aem-08}. It was also calculated for the excited state in free compact boson \cite{Capizzi:2020jed}. Here, we derived thermal corrections for the FCS for any two dimensional conformal field theory.
	We define the quantity:
	\be 
	g_{n}(Q_A,T)=\log f_{n}(Q_A,T)=ng\mathcal{F}_{n}(Q_A)e^{-2 \pi \Delta_{\psi} \beta /L}.
	\ee
	where $	f_{n}(Q_A,T)=\frac{{\mathcal{Z}}_n^{\text{(th)}}(Q_A)}{{\mathcal{Z}}_n^{(0)}(Q_A) }$. By choosing $n=1$, we find 
	\be 
	g_{1}(Q_A,T)=\left(\frac{\mathcal{X}_{1}(Q_A)}{{\mathcal{Z}}_1^{(0)}(Q_A)}-1 \right)g e^{-2 \pi \Delta_{\psi} \beta /L}.
	\ee
	This quantity represent the probability fluctuations. It expresses that, at low temperature, the ratio of the probability of finding a charge $Q_{A}$ at inverse temperature $\beta$ to zero one. It scales as $e^{-2 \pi \Delta_{\psi} \beta /L}$ with a coefficient that depends on the degeneracy of the first excited state, charge of the sector, and field content of the theory.
	
	\emph{ Entanglement measures--}	
	In general, the thermal correction to the symmetry-resolved R\'enyi and entanglement entropies take the following forms:
	\begin{widetext}
		\begin{eqnarray} 
		\label{dSn-corr}
		\delta S_n(Q_A) &=& \frac{ng}{1-n} \left[\frac{1}{n^{2\Delta_{\psi}}} \frac{ \sin^{2\Delta_{\psi}} \left( \pi x \right) }{\sin^{2\Delta_{\psi}} \left( \frac{ \pi  x}{n} \right)}\frac{\mathcal{X}_{n}(Q_A)}{{\mathcal{Z}}_n^{(0)}(Q_A)} -\frac{\mathcal{X}_{1}(Q_A)}{{\mathcal{Z}}_1^{(0)}(Q_A)}\right] 
		e^{-2 \pi \Delta_{\psi} \beta / L} +\nonumber \\
		&& o(e^{-2 \pi \Delta_{\psi} \beta / L}) 
		\  , \\
		\label{dSE-corr}
		\delta S_E(Q_A) &=& g \left[2 \Delta_{\psi}\left(1 - \pi x \cot \left( \pi x\right)\right)\frac{\mathcal{X}_{1}(Q_A)}{{\mathcal{Z}}_1^{(0)}(Q_A)}
		+ \partial_{n} \left(\frac{\mathcal{X}_{n}(Q_A)}{{\mathcal{Z}}_n^{(0)}(Q_A)}\right)\mid_{n=1}\right] e^{-2 \pi \Delta_{\psi} \beta/ L} \nonumber \\
		&& + o(e^{-2 \pi \Delta_{\psi} \beta / L})\ .
		\end{eqnarray}
	\end{widetext}
	Note that in the above expression we assume that the finite size of the system induces a mass gap that separates the ground state from the first excited state. Ae we see, these symmetry sector corrections  scales as $e^{-2 \pi \Delta_{\psi} \beta /L}$. Their coefficients, besides the size of the mass gap and degeneracy of the first excited state, depending on the four-point correlation function of primary fields.
	Compared to the total entanglement entropy, the scaling of these sectors is similar to the scaling of the total entanglement entropy, except that the correction scaling coefficients depend only on the mass gap and the degeneracy of the excited state and are independent of the charge of the sector\cite{Cardy:2014jwa}.
	
	A similar calculation was performed in ref.\  \cite{Capizzi:2020jed,Capizzi:2021zga}, where, the authors consider 
	symmetry-resolved entanglement and relative entropy in excited states of the CFT, by inserting two operators $\psi(z, \bar z)$ on all of the $n$ sheets of the cylinder. By contrast, the leading correction term in our calculation comes from the four-point function. 
	
	\emph{Examples.---}
	As a example, we specialize to compactified massless bosons with $c=1$~\cite{Di-MS, Muss, giam-b}. In this CFT, there are two holomorphic primary fields, vertex operator $V_{\beta}=e^{i \tilde{\beta} \phi_j}$ and derivative operator $i\partial \phi$, with scaling dimensions $h_{V}=K\frac{\tilde{\beta}^{2}}{2}$ and $h_{i\partial \phi}= 1$, respectively. The Luttinger parameter $K$ is related to the compactification radius via the bosonization relation. The lowest scaling primary field depends on range $\tilde{\beta}$ that is determined via Luttinger parameter $K$. In the following, we will consider both cases. 
	The generation of the Aharanov-Bohm flux is implemented by inserting the vertex operator $\mathcal{V}=e^{i \frac{\alpha}{2 \pi} \phi}$, with the scaling dimension 
	$h_\mathcal{V}=\bar{h}_\mathcal{V} =\frac{1}{2}\left( \frac{\alpha}{2\pi} \right)^2K$,
	such that generating the total phase $\alpha$ for the field upon going through the entire multi-sheeted Riemann surface $\mathcal{R}_n$\cite{Goldstein:2017bua}.
	If the excited state generated by the vertex operator $V_{\tilde{\beta}}=e^{i \tilde{\beta} \phi_j}$, we find that
	$G_{n,\alpha}(z,\bar{z})=
	e^{-iK\frac{\alpha \tilde{\beta} x}{n}}$
	It follows that, the exceed-cumulant generating function is $g_{n}(\alpha,T,x)=-iK\alpha \tilde{\beta} x/n$ ), which is universal. 
	\be\label{CMM4}
	\Delta C_{1,m}=g(-K\tilde{\beta}x)^{m}e^{-K \pi \tilde{\beta}^{2} \beta /L}.
	\ee
	The fluctuations  (and all the other cumulants) are derived by putting $n=1$. For example,
	Eq.~(\ref{CMM4}) with $n=1,m=2$ and $\Delta Q_A = Q_A - \langle  \hat{Q}_A \rangle$, implies that the excced in the variance in the conserved charge(number of particles) is  $(\Delta Q_A)^{2}=g(K\tilde{\beta}x)^{2}e^{-K \pi \tilde{\beta}^{2} \beta /L}$.
	If the excited state is induced by the derivative primary operator $i \partial \phi$, as a generator of low-dimensional primary state, the conformal block is 
	\be
	G_{n,\alpha}(z,\bar{z})=1-K\left( \frac{\alpha}{ \pi}\right)^{2}\sin ^{2}\left( \frac{\pi x}{n}\right),
	\ee
	The only non-zero exceed- cumulant is second cumulants, which is specify the charge fluctuation:
	\be
	\Delta C_{n,2}=\frac{2gK}{n} \frac{\sin ^{2}\left(\pi x\right)}{\pi^{2}}e^{-2 \pi \beta /L},
	\ee
	The variance excess is universal for any $n$. The physical meaning for $n=1$ is fluctuations of the charge.  Hence, we have
	\bea
	\frac{\mathcal{X}_{n}(Q_A)}{{\mathcal{Z}}_n^{(0)}(Q_A)}=1-\frac{ K\sin^{2}\left( \frac{\pi x }{n}\right) }{ \pi^{2} \sigma^{2}_{n}}\left(1-\left(\frac{Q_{A}}{\sigma_{n}} \right) ^{2} \right)
	\eea
	where $\sigma_{n}^{2}=\frac{K\log \left(l \right) }{n\pi^{2}}$.  
	We find the universal correction for the full counting statics (FCS) as
	\be 
	\label{CORR-1}
	g_{1}(Q_{A})=\frac{K\sin^{2}\left( \pi x\right)  }{\pi^{2}\sigma_{1}^{2}}\left( \left(\frac{Q_{A}}{\sigma_{1}} \right) ^{2}-1\right) e^{-2 \pi \beta /L}.
	\ee
	It denotes the probability fluctuations.
	The resolved-symmetry entanglement entropy takes the following form:
	\begin{widetext}
		\bea
		&&	\delta S_E(Q_A)=	\delta S_E+\mathcal{B}(Q_A)e^{-2 \pi  \beta/ L}
		,
		\cr \nonumber\\ \qquad
		&&
		\mathcal{B}(Q_A)= 
		\left(3\sin^{2}\left( \pi x \right)-\pi(1+x)\sin\left( 2\pi x \right) \right)\frac{1}{\ln (l)}
		+\left(-4\pi\sin^{2}\left( \pi x \right)+\pi(1+x)\sin\left( 2\pi x \right) \right)\frac{\pi Q_A^{2}}{K\ln (l)^{2}} .
		\eea
	\end{widetext}
	where $	\delta S_E=2 \left(1 - \pi x \cot \left( \pi x\right)\right)$ \cite{Cardy:2014jwa}. We see that the thermal corrections of the charge-sector contributions, at order $\ln (l)^{-2}$, are charge-dependent. The same result for excited state associated for the derivative operator is derived in \cite{Capizzi:2020jed}, the scaling of the terms are the same and the equipartition of the entanglement breaks at order $\ln (l)^{-2}$, but the coefficient of those terms are cumbersome. Here the coefficients are simple.
	
	\emph{Discussion.---}
	In this work, we have derived a formula for thermal corrections to the symmetry-resolved R\'{e}nyi and entanglement entropies for general two-dimensional conformal field theories on a circle. Besides the size of the mass gap and the degeneracy of the first excited state, these terms depend only on the four-point function of primary fields.
	It is worth noting that, until now, these thermal corrections have only been studied in the free case, whereas we found it for any two-dimensional CFT. Specially, we have derived the thermal corrections to full counting statistics, and the excess-cumulant generating function.
	We also have defined the probability fluctuations functions. It expresses that, at low temperature, the ratio of the probability of finding a charge $Q_{A}$ at inverse temperature $\beta$ to zero one. It scales as $e^{-2 \pi \Delta_{\psi} \beta /L}$ with a coefficient that depends on the degeneracy of the first excited state, charge of the sector, and field content of the theory.
	We have explicitly evaluated thermal corrections for the entanglement entropy and FCS in the free compact boson theory for both derivative and vertex operators.  
	We have found that in the first case, the entanglement equipartition break at order of $\ln (l)^{-2}$.

	\begin{acknowledgments}
		\emph{Acknowledgements.---}
		Thanks to Pasquale Calabrese, Luca Capizzi, Ghadir Jafari, Mojtaba Mohammadi, Sepideh Mohammadi, Sara Murciano, Ali Naseh, Reza Pirmoradian, Hesam Soltanpanahi, and Behrad Taghavi for helpful discussions on our manuscript.

	\end{acknowledgments}
	
	\section{Supplemental Material}\label{appendix1}
	In this section we will derive the expression (\ref{EQ-M1}).
	The state/operator correspondence associates to each state $|\psi \rangle$ a local operator $\psi(x,t)$ such that $|\psi \rangle=\lim_{t \to -\infty} \psi(x,t) |0 \rangle$ and similarly $\langle \psi | = \lim_{t \to \infty} \langle 0 | \psi(x,t)$. 
	On the other hand, using/by the path integral formalism, the trace over an $n$th power of the reduced density matrix Tr$(\rho_{A})^{n}$ reduces to a partition function on an $n$-sheeted Riemann surface $\mathcal{R}_{n}$, the multisheeted cylinder, branched over the interval $A$.
	The expression (\ref{M-EQ4}) is then the four-point function on this multisheeted cylinder.
	
	By the Riemann-Hurwitz theorem, this multisheeted Riemann surface $\mathcal{R}_{n}$ has genus zero, and can be uniformized  through the conformal map\cite{Cardy:2014jwa}
	\be
	\zeta^{(n)} = \left( \frac{e^{2 \pi i w/L} -  e^{i \theta_2}}{e^{2 \pi i w/L} - e^{i \theta_1}} \right)^{1/n}  \ .
	\ee
	This map takes the multisheeted cylinder to the plane. The parameters $\theta_1$ and $\theta_2$ are selected so that the map is branched over the interval $A$ with endpoints on the cylinder, such that
	$
	\theta_2 - \theta_1 = \frac{2 \pi \ell}{ L}
	$.  
	Subsequently, the insertion points of the primary fields on the $j$th cylinder at the points  $t = -\infty$ and $t = \infty$ are transformed to the points $\zeta^{(n)}_{4}\equiv\zeta^{(n)}_{-\infty} = e^{i(\theta_2 - \theta_1)/n + 2 \pi i j/ n}$ and $\zeta^{(n)}_{3}\equiv\zeta^{(n)}_\infty = e^{2 \pi i j/n}$, respectively, on the $\zeta^{(n)}$ plane. While the operators on the end points of the interval are mapped to the points $\zeta^{(n)}_{1}=\infty$ and $\zeta^{(n)}_{2}=0$ on the $\zeta^{(n)}$ plane.  
	Using the following identity
	\be
	\frac{ d \zeta^{(n)} / d w}{d \zeta^{(1)} / dw} = \frac{1}{n} \frac{\zeta^{(n)}}{\zeta^{(1)}} \ .
	\ee
	we find that the transformed four-point function is 
	\begin{widetext}
		\begin{align}
		\mathcal{M}_{n}(\alpha) = & \frac{1}{n^{2\Delta_{\psi}}}\left(\frac{\zeta^{(n)}_{3}\zeta^{(n)}_{4}}{\zeta^{(1)}_{3}\zeta^{(1)}_{4}}\right)^{h_{\psi}}\left(\frac{\bar{\zeta}^{(n)}_{3}\bar{\zeta}^{(n)}_{4}}{\bar{\zeta}^{(1)}_{3}\bar{\zeta}^{(1)}_{4}}\right)^{\bar{h}_{\psi}}
		\frac{\left\langle \mathcal{V_{\alpha}}\left(\zeta^{(n)}_{1},\bar{\zeta}^{(n)}_{1}\right)\mathcal{V_{-\alpha}}\left(\zeta^{(n)}_{2},\bar{\zeta}^{(n)}_{2}\right)\psi \left(\zeta^{(n)}_{3},\bar{\zeta}^{(n)}_{3}\right)\psi \left(\zeta^{(n)}_{4},\bar{\zeta}^{(n)}_{4}\right)\right\rangle}{\left\langle \mathcal{V_{\alpha}}\left(\zeta^{(n)}_{1},\bar{\zeta}^{(n)}_{1}\right)\mathcal{V_{-\alpha}}\left(\zeta^{(n)}_{2},\bar{\zeta}^{(n)}_{2}\right)\right\rangle  \left\langle \psi \left(\zeta^{(1)}_{3},\bar{\zeta}^{(1)}_{3}\right)\psi \left(\zeta^{(1)}_{4},\bar{\zeta}^{(1)}_{4}\right)\right\rangle}.
		\end{align}
	\end{widetext}
	The above expression can be simplified such that we reach to the expression (\ref{EQ-M1}).


\end{document}